# A Spin Glass Model of Human Logic Systems

**Fariel Shafee**
Physics Department
Princeton University
Princeton, NJ 08540

---

In this paper, we model logic networks interacting among themselves and also with the environment.

We assume that although each agent is rational in the sense that he tries to maximize her utilities, the utilities themselves are not absolute and depend in part on slightly varying genetic make-ups leading to innate differences in preferences, varying acquired tastes depending on local non homogeneity among interacting agents and the environment, and also slight differences in perception and vision.

The agents acquire knowledge by interacting with nature and other agents. The preferences serve as the "truths taken for granted" or axioms for an agent's logic system and he uses the knowledge to enhance his utilities.

We can assume perception as an agent's connecting or interacting points with the environment and other agents. However, if we model an agent's perception of the environment as a shadow of a higher dimensional manifold, slight variations in the projection will lead to differing ideas about the environment. An agent will interact with the environment based on the projection he has knowledge of, and the interaction will have effects on another agent's projection, which may not coincide with the first projection. Hence, although both agents act on the "truth" they perceive, one's actions might have conflicting results on another agent's universe.

Also, agents' tastes or preferences dictate to some extent how they want to "change" the environment, or what their visions of the future are. One agent might have an inclination toward one of many possible states of nature while the second agent might prefer another. However, since both agents must interact with the same environment they are connected to, conflicts arise from different visions about the future associated with the same environment.

This situation can be modeled in a way similar to the many-universe hypothesis [1]. We assume that the environment is "collapsing to a future." The collapse can occur to one of many superposed states and the agents bid on different states. A similar model with quantum collapse with bidding has been proposed [2, 3].

The bidding also involves an agent's calculation of risk factors associated with investing his time and energy in a certain "stock." Since each agent has only a finite amount of time and energy, one can calculate upto a finite number of leading terms in the many variables associated when calculating risks. Each agent decides where to truncate the calculation, depending on his weighed preferences.

We connect the logic networks [4] in a model akin to the spin model [5]. Here, we can draw similarities with thermodynamics where a micro-system (in this case the perceptory organs - coupled to the neural network) is placed in conjunction with a macro-system (which is the environment). These logic networks placed in a society may be visualized as a small thermodynamic spin lattice where the axioms (i, j) residing in different networks are coupled to one another by coupling constants $J_{ij}$, and this lattice itself is coupled to a bigger lattice, which is the environment. However, since we can define the environment lattice to be huge compared to the neural network lattice, we can take average values for interaction purposes and couple the lattice with the neural network lattice with some multidimensional coupling factor.

In each of the lattices, spins are at a quantum level described as "states" which can coexist in many orthonormal superpositions. However, when the smaller lattice interacts with the bigger lattice, the coupling causes the environment lattice to collapse to a certain value. This value will depend on the probabilistic coefficients of the wave functions and most of the time it would yield the expected value. So an average person will end up with an average set of axioms. Each of the agents has a certain set of axioms to start with. Again, these agents are coupled with one another in a lattice. We argue that the agents are not connected with one another with a random coupling constant $J_{ij}$, but some rules are defined, and also that these $J_{ij}$s are updatable according to the specific state of the entire network.

1. $J_{ij}$ is not symmetric, i.e., $J_{ij} \neq J_{ji}$. The value of $J_{ij}$ depends on i possessing axioms that necessitate the existence of j. So i will be coupled to j more strongly if i possesses axioms that require the existence of j. Now each agent will have the following behaviors in the game:

1. Each agent i will tend to change its neighbors' axioms if the neighbors' axioms contain contradictions of its axioms. The frequency and strength of flipping would depend on a coupling constant $C_{ij}$. $C_{ij}$ depends on the following:

a. The evolution of the logical code developed in i that contains that particular axiom, i.e. the networking of the certain axiom in the logic network of i. More accurately, the number of decisions produced by i that reflect the use of the particular axiom. In other words, i will tend to flip a neighbors anti-axiom with more effort if the axiom has become an important part of its network, and any future attempt of j's flipping it would cost i dearly.

b. The determination of the number of the contradictory axiom in the neighboring agents' logic network. An increased frequency of anti-axioms in is network would increase the possibility of an anti-axiom to be used in a future decision.

c. The effect of j's decision on its environment (might be caused by physical distance between the two agents). The total strength of coupling between i and j (i → j ) would be - $J^n_{ij}$ (state$^n_i$ → state$^n_j$) + $C^m_{ij}$ (state$^m_i$ → state$^m_j$), where n and m are states or registers representing axioms.

We can implement this scheme by linking two agents with appropriate logic gates. The value of this coupling could be described as feelings of agent i towards agent j. This total coupling will, at a macroscopic level, cause i to play for or against j, i.e. collaborate with j or work against the existence of j.

2. An axiom will flip if the effect of the neighbor's having the same axiom state causes the coupling to exceed a flipping energy. The flipping energy depends on certain axioms' connectivity with other axioms in the agent's cognitive network.

3. All agents must possess an axiom we shall call self preservation or preservation of the network in random probability. We label this axiom P. Agents containing neither of these axioms cannot contribute to the existence of the agent or to the network. In that case those agents' axioms, if few, will be flipped by other agents; or they will self-destroy. A later paper will discuss the effect of agents possessing destructive axioms in their cognitive network, or any critical number that will bound the fraction of agents with self destructive axioms in a network. However, in this paper, we assume that all agents possess P.

The probability of obtaining a certain state of an axiom by another agent can now be written down in a simplified version by the formula:

$$F(A) = \sum C_{ijA} f_1(\text{flip}) - \sum J_{ijA} f_2(\text{stabilize}) - f_3(1/R) + K <A>$$

Here, A is the axiom, R is a resistance factor for flipping the axiom depending on how entangled the axiom is in the agents own cognitive network, K is the coupling of the agent with nature, and $<A>$ is the expectation probability of the axiom in nature. We can see that this formula is very similar to the formula for a classical neural network, except that the coupling constants, unlike the weight factors in a regular neural network, do not sum up to 1. Also, instead of adding a term in $<A>$, it might be more realistic to add a term $F(<A>)$. Here F(A) is a switching function that takes on a value of either -1 or 1, depending on whether the RHS exceeds a certain threshold. So every time A or Ã (not-A) flips, the C's and the J's are interchanged, and R is updated to a new value that needs to be updated with the accumulation of new axioms that get entangled with the flipped clause.

A cost of entropy term must be added depending on whether the agent has an axiom involving taking entropy into account or an axiom that ignores entropy in its network. Entropy has been defined variously in different contexts, usually as a measure of information needed about the state of the system, which can be related to the complexity of the system [6, 7], whereas information itself has been related to symmetry-breaking [8]. In a macro-system the dynamics guides the system to a state of least free energy, which is related to energy (U) and the entropy (S) by

$$F = U - TS$$

T being the temperature.

In condensed matter physics T is found to be analogous to inverse time, and hence entropy behaves like energyX time, which is a reasonable guide to the introduction of this quantity, because it is necessary to reduce the utility (F) by a quantity which involves the complexity of the system that the agents have to unravel to evaluate the different terms from the interactions of the axioms, and also the inverse time (rate or speed) involved in the process. This opens up the possibility of highly intricate games. The nearly chaotic nature of even small world examples with variable spin-like constituents in simulation examples have been reported recently [9], but the role of entropy, or of ignoring its role by a class of agents using an "ignore entropy axiom" remains to be studied thoroughly.